\documentclass[preprint,showpacs,preprintnumbers,amsmath,amssymb]{revtex4}

\usepackage{graphicx}
\usepackage{dcolumn}
\usepackage{bm}
\usepackage[dvips]{color}

\begin{document}

\title{Hydrogen Dissociation and Diffusion on Transition Metal(=Ti,Zr,V,Fe,Ru,Co,Rh,Ni,Pd,Cu,Ag)-doped Mg(0001) Surfaces}

\author{M. Pozzo$^{1,2}$}
\author{D. Alf\`{e}$^{1,2,3,4}$}%
\email{d.alfe@ucl.ac.uk} 
\affiliation{
$^1$Material Simulation Laboratory, University College London, 
Gower Street, London WC1E 6BT, United Kingdom \\
$^2$Department of Earth Sciences, University College London, Gower
Street, London WC1E 6BT, United Kingdom \\
$^3$Department of Physics and Astronomy, University College London, Gower
Street, London WC1E 6BT, United Kingdom \\
$^4$London Centre for Nanotechnology, University College London, 17-19 Gordon
Street, London WC1H 0AH, United Kingdom}%

\date{\today}

\begin{abstract}
The kinetics of hydrogen absorption by magnesium bulk is affected by
two main activated processes: the dissociation of the H$_2$ molecule
and the diffusion of atomic H into the bulk. In order to have fast
absorption kinetics both activated processed need to have a low
barrier.  Here we report a systematic ab-initio density functional
theory investigation of H$_2$ dissociation and subsequent atomic H
diffusion on TM(=Ti,V,Zr,Fe,Ru,Co,Rh,Ni,Pd,Cu,Ag)-doped Mg(0001)
surfaces. The calculations show that doping the surface with TM's on
the left of the periodic table eliminates the barrier for the
dissociation of the molecule, but the H atoms bind very
strongly to the TM, therefore hindering diffusion. Conversely, TM's on
the right of the periodic table don't bind H, however, they do not
reduce the barrier to dissociate H$_2$ significantly.  Our results
show that Fe, Ni and Rh, and to some extent Co and Pd, are all exceptions,
combining low activation barriers for both processes, with Ni being the best
possible choice.
\end{abstract}



\maketitle

\section{Introduction}\label{intro}

Hydrogen is regarded by many as a possible energy vehicle (or fuel)
for future mobile applications, and has been targeted
to replace the current use of liquid hydrocarbons in the next few
decades~\cite{schlapzutt01}.  Unlike fossil fuels, it is an
environmentally friendly, non-polluting fuel simply because its
combustion product is water, providing it is produced by renewable
energy resources, obviously.

One of the main challenges faced by the development of the so called {\em
hydrogen economy} is the capability of storing hydrogen safely
and efficiently. For mobile applications the storage materials need to
satisfy a number of requirements: i) they need to be capable of 
storing hydrogen in excess of 6.5\% in weight, ii) the kinetics of 
absorption has to be fast, i.e. with time scales of minutes, and 
iii) the temperature at which they release hydrogen (the decomposition 
temperature) needs to be ideally in the range 20-100$^o$C. Cyclability 
of the material is also a desired property.

Metal hydrides are natural hydrogen storing materials, and the
relatively strong bonds between hydrogen and the host metal satisfy
the safety requirement. Unfortunately, however, no material which has
all the properties mentioned above exists today (see for example the
review by Sakintuna et al.~\cite{sakintuna07}).

Magnesium hydride, MgH$_2$, satisfies some of the above
requirements. It has high storage capacity (7.6 wt \%), good
cyclability and it is relatively inexpensive. However, its enthalpy of
formation is too high (-76 kJ/mol), requiring temperatures in excess
of 300$^o$C to decompose it into H$_2$ and Mg bulk. The formation of
the hydride also has slow kinetics~\cite{huot99,zaluska01}, making
this material not good enough. However, MgH$_2$ represents a good test
material to study how various treatments can affect its properties. In
particular, it has been found that by doping the material with
transition metals can weaken the Mg-H bond and reduce the stability of
the hydride (see for example
\cite{zaluska99,bobet02,song04,shang04,vegge05,bhat07}, and references
therein). Ball-milling further enhance the sorption processes by
increasing the number of possible paths for the diffusion of H (see
for example \cite{zaluski95,zaluska99,shang04} and references
therein). A new method of chemical fluid deposition in supercritical
fluids (SCF) has also recently been used on metal hydrides
\cite{bobet07}. This method offers the same sorption properties of
ball milled samples, but with a hugely improved cyclability (the
catalytic effect of the metal being almost constant for SCF samples,
while decreasing for ball-milled samples after about 100 cycles) and
in particular shows that the catalytic effect of Ni on hydrogen
sorption processes is higher than that of Pd.

Liang et al.~\cite{liang99} and Schulz, Liang \& Huot~\cite{schultz01}
found that V and Ti are better catalysts than Ni for hydrogen
absorption and desorption from MgH$_2$-metal composites, showing
faster absorption kinetics at T $\sim$~300~$^o$C and faster desorption
kinetics above $\sim$~250~$^o$C than other 3$d$-elements investigated.
They also found an enthalpy of hydride formation for different
catalysts similar to that of MgH$_2$. By contrast, theoretical
calculations and experimental results of Song et al.~\cite{song04} and
Shang et al.~\cite{shang04} led to different conclusions. They found
that the stability of MgH$_2$-Ni is reduced when compared to that of
MgH$_2$-Ti. Moreover (and contrary to the experimental findings of
Liang et al.~\cite{liang99}), the heat of formation of the metal-doped
MgH$_2$ hydrides is smaller than that of MgH$_2$. In particular,
MgNiH$_2$ shows a smaller enthalpy of formation than MgTiH$_2$.

Zaluska et al.~\cite{zaluska99} used Li, Al, V, Mn, Zr and Y as
catalysts for the hydrogenation/dehydrogenation of Mg alloy
samples. According to their results, V remarkably improves the H
absorption kinetics, but Zr is better for lower temperature
H. However, the best kinetic results are achieved with mixtures,
i.e. V + Zr or Mn + Zr Mg alloys. Bobet et al.~\cite{bobet00} have
shown that the hydrogen storage properties are enhanced when using
reactive mechanical alloying of Mg + 10 wt.\% Co, Ni and Fe
mixtures. Co, unlike Ni, is found to significantly increase the
quantity of MgH$_2$ formed. However, Bobet et al.~\cite{bobet02} later
reported that the hydrogen sorption properties of Mg-Co mixtures are
less effective than those reported for MgH$_2$-metal mixtures.
Gutfleish et al.~\cite{gutf05} have recently presented results
achieved with a Mg sample alloyed with Ni (1 wt \%) and Pd (0.2 wt
\%). Their sample shows excellent hydrogen absorption/desorption
kinetics and cyclic stability, exhibiting an overall reversible H$_2$
storage capacity of 6.3 wt. \%.

Previous theoretical and experimental investigations over pure
surfaces of transition metals belonging to the left of the periodic
table have shown that H$_2$ dissociation is promoted, but also that
the bonding between the hydrogen atoms and the metal is strong
(see~\cite{ward80} and references therein; see
also~\cite{christm88,hamnorNAT95}).

In our previous paper~\cite{pozzo} we have shown that H$_2$
dissociation on the metal(Ni,Ti)-doped Mg surface has a barrier
similar to that on the corresponding pure metal (111) surface, and
also that the strength of the hydrogen-metal bond is similar to that
on the pure metal surface.  The strength of the H-TM (TM=Ni,Ti)
bonding was found to be correlated to the height of the diffusion
barrier. We therefore might expect to see an analogous trend in the
dissociation and diffusion barriers by doping the Mg(0001) surface
with various transition metals. In fact, we will show that the
elements on the left of the periodic table make the H$_2$ dissociation
barrier to vanish but are responsible for high diffusion barriers,
while those on the right cannot catalyse the dissociation of the
molecule. Among the elements studied here, we found that Fe, Ni and
Rh, and to some extent Co and Pd offer a good compromise between the
promotion of dissociation and the hindering of diffusion, and qualify
as good catalysts for accelerating the kinetics of hydrogen
absorption.

\section{Computational method}\label{methods}

All the DFT calculations were performed with the ab-initio simulation
package VASP~\cite{vasp} using the projector augmented wave (PAW)
method~\cite{blochl94,kresse99} and the PBE exchange-correlation
functional~\cite{pbe}. An efficient charge density extrapolation was
used to speed up the calculations~\cite{alfe99}. We used a plane-wave
basis set to expand the electronic wave-functions with the same
plane-wave energy cut-off of 270 eV as in~\cite{pozzo}, which
guarantees convergence of adsorption energies within 1 meV.  Surfaces
were modeled using periodic slabs, with 5 atomic layers and a vacuum
thickness of 10~\AA. The topmost three atomic layers were allowed to
relax, while the bottom two were held fixed to the positions of bulk
Mg. Calculations were performed using 2x2 surface unit cells, with
9x9x1 {\bf k}-point grids and replacing one of the four surface Mg
atoms by one TM(=Ti,V,Zr,Fe,Ru,Co,Rh,Ni,Pd,Cu,Ag) atom.  These
settings were extensively tested and guarantee convergence of
activation energies to better than 0.02 eV.  Activation energies have
been calculated with the nudged elastic band (NEB) method~\cite{neb}
using 17 replicas, which proved to be sufficient to reach convergence
of activation energies to better than 0.01 eV, and display all the
main features of the minimum energy path. The initial state of the NEB
calculations for the dissociation of H$_2$ is represented by the
hydrogen molecule sitting on top the TM at a distance of 5~\AA, and
the final state is the most energetically favourable among four
possible adsorption sites for the two dissociated hydrogen atoms (see
Ref.~\cite{pozzo} for details). For the diffusion process, the initial
state is represented by the final state of the dissociation process,
and the final state by a configuration where one of the two hydrogen
atoms has been displaced into a nearby hollow site (see details in
Section~\ref{H2diss}).

\begin{figure*}
\rotatebox{0}{\scalebox{0.95}[0.45]{\includegraphics{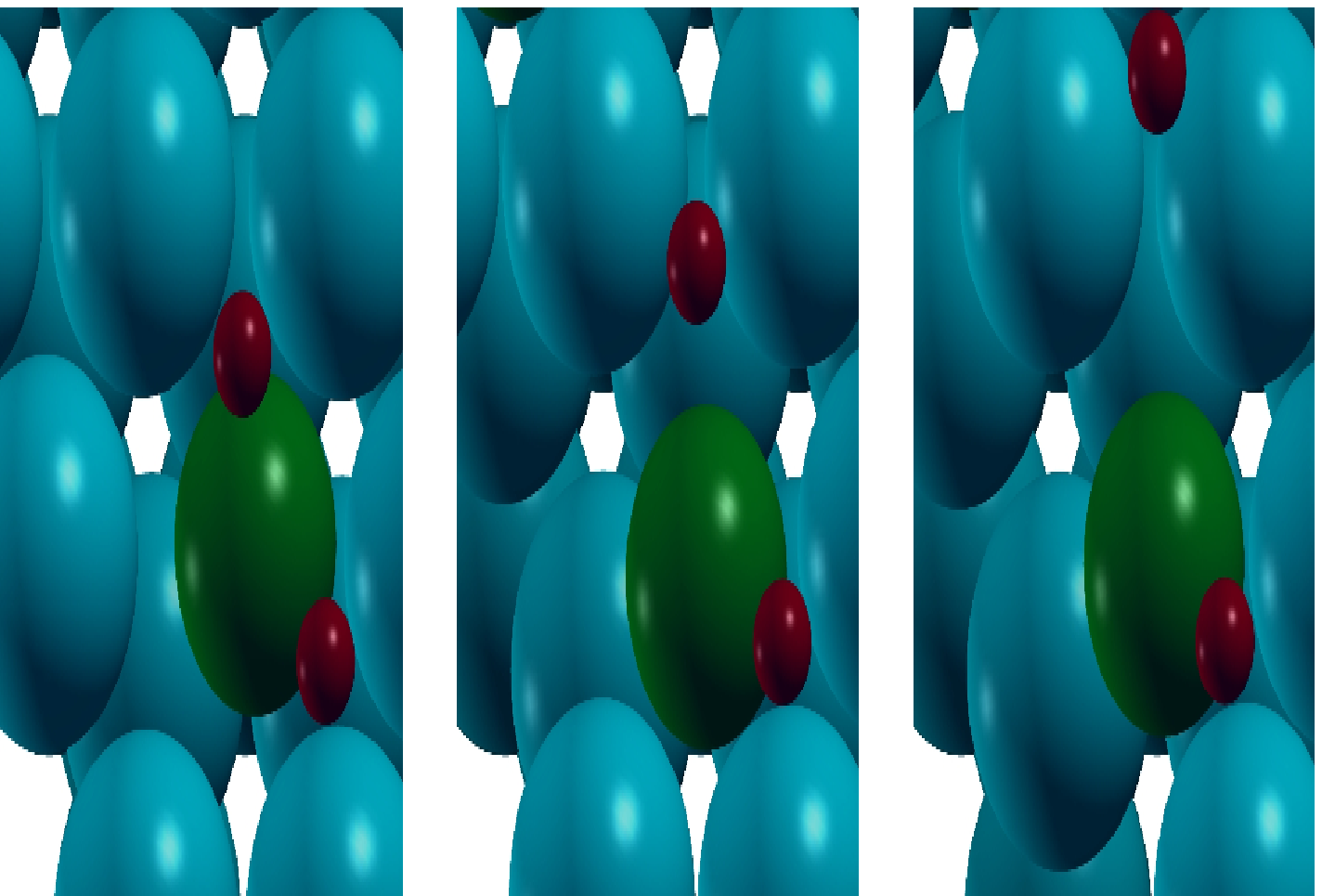}}}
\caption{\label{MgFeHdiff}}
\phantom{xxxxxxxxxxxxxxxxxxxxxxxxxxxxxxxxxxxxxxxxxxxxxxxxxxxxxxxxxxxxxxxxxxxxxxxx} 
\phantom{xxxxxxxxxxxxxxxxxxxxxxxxxxxxxxxxxxxxxxxxxxxxxxxxxxxxxxxxxxxxxxxxxxxxxxxx}
\phantom{xxxxxxxxxxxxxxxxxxxxxxxxxxxxxxxxxxxxxxxxxxxxxxxxxxxxxxxxxxxxxxxxxxxxxxxx} 
\phantom{xxxxxxxxxxxxxxxxxxxx} 
\phantom{xxxxxxxxxxxxxxxxxxxx} 
\phantom{xxxxxxxxxxxxxxxxxxxx} 
\phantom{xxxxxxxxxxxxxxxxxxxx} 
\end{figure*}

Figs.~\ref{MgAgH2diss} and~\ref{MgFeHdiff} have been made using the
XCRYSDEN software~\cite{xcrysden}.

\section{Results}\label{results}

In the following section we report calculations for the bulk
structural parameters of the various elements investigated, as a test
of the quality of the PAW and the PBE exchange-correlation
functionals.  In Sec.~\ref{H2diss} we report results for the H$_2$
dissociation and diffusion barriers, which we also analyse in terms of
the position of the centre of the $d$-band of the various transition
metals employed as dopants on the Mg(0001) surface.

\begin{table}
\footnotesize{
\caption{\label{table1} Bulk properties of pure transition metals 
(TM$=$Ti,V,Fe,Co,Ni,Cu,Zr,Ru,Rh,Pd,Ag). For each element we report
the bulk lattice constant $a$ (together with $c/a$ for hcp metals), the bulk modulus k$_0$,
the electrons treated as valence (VE) and the core radius r$_{core}$ of the PAW potentials.
References for values previously reported in literature follow in the last column.}
\begin{ruledtabular}
\begin{tabular}{lcccl}
 & $a$ (\AA), $c/a$ & k$_0$ (GPa) & VE, r$_{core}$ (\AA) & Reference \\
\hline
Ti      & 2.92, 1.583 & 120 & 3d$^2$4s$^2$, 1.5 & Ref.~\onlinecite{pozzo} \\
\vspace{-1.1cm}\\
        & 2.92, 1.583 & 118 &             & Ref.~\onlinecite{jahna05} \\
\vspace{-1.1cm}\\
$[$Expt.$]$ & [2.95, 1.59] & [105]  & & Ref.~\onlinecite{kittel} \\
V       & 3.00 & 179 &3s$^2$3p$^6$3d$^3$4s$^2$, 1.2 & This work\footnotemark[1] \\ 
\vspace{-1.1cm}\\
        & 3.00 & 182 &  & Ref.~\onlinecite{vermod07} \\ 
\vspace{-1.1cm}\\
        & 2.99 & 185 &  & Ref.~\onlinecite{jahna05} \\
\vspace{-1.1cm}\\
$[$Expt.$]$ & [3.00] & [162] &  & Ref.~\onlinecite{kittel} \\
Fe     & 2.84 & 169 & 3d$^6$4s$^2$, 1.2 & This work\footnotemark[1] \\
\vspace{-1.1cm}\\
       & 2.71 & 281 &  & Ref.~\onlinecite{mehpap96}\footnotemark[3] \\
\vspace{-1.1cm}\\
$[$Expt.$]$ & [2.87] & [168] &  & Ref.~\onlinecite{kittel} \\
Co     & 2.49, 1.617 & 212 & 3d$^7$4s$^2$, 1.2 & This work\footnotemark[1] \\
\vspace{-1.1cm}\\
       & 2.40, 1.62 & 384 &  & Ref.~\onlinecite{mehpap96}\footnotemark[3] \\
\vspace{-1.1cm}\\
$[$Expt.$]$ & [2.51, 1.62] & [191]  &  & Ref.~\onlinecite{kittel} \\
Ni      & 3.52 & 194 & 3d$^8$4s$^2$, 1.2 & Ref.~\onlinecite{pozzo} \\
\vspace{-1.1cm}\\
        & 3.52 & 194            &  & Ref.~\onlinecite{kresshafn00}  \\
\vspace{-1.1cm}\\
        & 3.52 & 201            &  & Ref.~\onlinecite{pozzo07} \\
\vspace{-1.1cm}\\
$[$Expt.$]$ & [3.52] & [186] &  & Ref.~\onlinecite{kittel} \\
Cu      & 3.64 & 136 & 3d$^{10}$4s$^1$, 1.2 & This work\footnotemark[1] \\
\vspace{-1.1cm}\\
        & 3.63 & 142 &  & Ref.~\onlinecite{dasilva06} \\ 
\vspace{-1.1cm}\\
$[$Expt.$]$  & [3.61] & [137] &  & Ref.~\onlinecite{kittel} \\
Zr      & 3.24, 1.602 & 93 & 4s$^2$4p$^6$4d$^2$5s$^2$, 1.3 & This work\footnotemark[1] \\
\vspace{-1.1cm}\\
        & 2.99, 1.86  & 108 &  & Ref.~\onlinecite{mehpap96}\footnotemark[2] \\
\vspace{-1.1cm}\\
        & 3.23, 1.600 & 101 &  & Ref.~\onlinecite{yaoyou07} \\
\vspace{-1.1cm}\\
$[$Expt.$]$  & [3.23, 1.59] & [83] &  & Ref.~\onlinecite{kittel} \\
\vspace{-1.1cm}\\
       & [3.23, 1.59]& [92 $\pm$ 3] &  & Ref.~\onlinecite{zhao05} \\
Ru     & 2.72, 1.578 & 310 & 4d$^7$5s$^1$, 1.4 & This work\footnotemark[1] \\
\vspace{-1.1cm}\\
       & 2.68, 1.59 & 360 &  & Ref.~\onlinecite{mehpap96}\footnotemark[3] \\
\vspace{-1.1cm}\\
       & 2.69, 1.606 & 322 &  & Ref.~\onlinecite{yao07} \\
\vspace{-1.1cm}\\
$[$Expt.$]$ & [2.71, 1.58] & [321]  & & Ref.~\onlinecite{kittel}  \\
Rh      & 3.84 & 251 & 4d$^8$5s$^1$, 1.3  & This work\footnotemark[1] \\
\vspace{-1.1cm}\\
        & 3.83 & 259 &  & Ref.~\onlinecite{ganduglia99} \\
\vspace{-1.1cm}\\
        & 3.86 & 258 &  & Ref.~\onlinecite{pozzo07} \\
\vspace{-1.1cm}\\
$[$Expt.$]$ & [3.80] & [270] &  & Ref.~\onlinecite{kittel} \\
Pd     & 3.95 & 163 & 4d$^{10}$, 1.3 & This work\footnotemark[1] \\
\vspace{-1.1cm}\\
       & 3.95 & 163 &  & Ref.~\onlinecite{dasilva06} \\
\vspace{-1.1cm}\\
$[$Expt.$]$ & [3.89] & [181] &  &  Ref.~\onlinecite{kittel} \\
Ag     & 4.17 & 88 & 4d$^{10}$5s$^1$, 1.3 & This work\footnotemark[1] \\
\vspace{-1.1cm}\\
       & 4.20 & 87 &  &  Ref.~\onlinecite{li02} \\
\vspace{-1.1cm}\\
$[$Expt.$]$ & [4.09] & [101] &  & Ref.~\onlinecite{kittel} 
\footnotetext[1]{Reported values do not include room temperature thermal expansion.}
\footnotetext[2]{From tight-binding calculations.}
\footnotetext[3]{From tight-binding non-magnetic calculations.}
\end{tabular}
\end{ruledtabular}
}
\end{table}

\subsection{Bulk parameters}

We obtained bulk structural properties of the pure transition metals
by calculating energy versus volume curves, and fitting them to a
Birch-Murnaghan equation of state~\cite{murna}. The elements
investigated here were: Zr, V, Fe, Ru, Co, Rh, Pd, Cu and Ag, together
with Ti and Ni already presented in Ref.~\cite{pozzo}.  The bulk
parameters were derived using 13x13x13 and 18x18x12 {\bf k}-point
grids for those metals with the cubic and the hexagonal structure
respectively.  The corresponding standard version of the PAW
functional was used for all of them, with the exception of V and Zr
for which we used the version of the PAW treating respectively the
3s$^2$3p$^6$3d$^3$4s$^2$ and 4s$^2$4p$^6$4d$^2$5s$^2$ electrons in
valence (which give results closer to the experimental values than the
standard versions of the PAW functionals). Calculated bulk parameters
values are reported in Table~\ref{table1}, together with the details
of the PAW potentials, which include the electrons treated in valence
and the core radii. Overall, the lattice parameter $a$ is always
overestimated, and the bulk modulus is underestimated with respect to
the experimental values. This is in agreement with the findings from
previous theoretical calculations, and is typical of the PBE
functional.  Among the elements investigated, only Ni, Co and Fe are
magnetic. We find a magnetic moment of 0.63~\cite{pozzo}, 1.70 and
2.15 $\mu_B$/atom for Ni, Co and Fe respectively, which are in
agreement with the corresponding experimental values of 0.61, 1.71 and
2.22 $\mu_B$/atom~\cite{kittel}. However, as discussed in the next
Section, Fe is the only element which required spin-polarised
calculations.

\begin{table}
\caption{\label{table2} Activation energies for H$_2$ dissociation (E$_{diss}$)
on the pure Mg and metal-doped Mg surfaces (ordered by increasing atomic number).}
\begin{ruledtabular}
\begin{tabular}{lc}
Metal surface & E$_{diss}$ (eV) \\
\hline
pure Mg  & 0.87\footnotemark[1], 0.4\footnotemark[2]$^,$\footnotemark[3], 
0.5\footnotemark[4]$^,$\footnotemark[5], 1.15\footnotemark[6], 1.05\footnotemark[7],
0.95\footnotemark[8] \\
$[$Expt.$]$ & 1.0\footnotemark[9], 0.75$\pm$0.15\footnotemark[10] \\ \\

Ti-doped Mg & null\footnotemark[1], negligible\footnotemark[7]\\

V-doped Mg & null\footnotemark[11]\\

Fe-doped Mg & 0.03\footnotemark[11]\\ 

Co-doped Mg & 0.03\footnotemark[11]\\

Ni-doped Mg & 0.06\footnotemark[1]\\ 

Cu-doped Mg & 0.56\footnotemark[11]\\

Zr-doped Mg & null\footnotemark[11]\\
 
Ru-doped Mg & null\footnotemark[11]\\

Rh-doped Mg & 0.04\footnotemark[11]\\ 

Pd-doped Mg & 0.39\footnotemark[11]\\ 

Ag-doped Mg & 1.18\footnotemark[11]\\ \\
\end{tabular}
\end{ruledtabular}
\footnotetext[1]{Ref.~\onlinecite{pozzo}.}
\footnotetext[2]{Ref.~\onlinecite{hjelmberg79} for a jellium system.}
\footnotetext[3]{Ref.~\onlinecite{bird93}, from DFT LDA calculations and PES. 
This lower value as compared to other calculations is explained as due to the well known 
LDA over-binding.}
\footnotetext[4]{Ref.~\onlinecite{johansson81} for a jellium system and PES.}
\footnotetext[5]{Ref.~\onlinecite{norskhoum81} for a jellium system and PES.}
\footnotetext[6]{Ref.~\onlinecite{vegge04} from DFT RPBE.}
\footnotetext[7]{Ref.~\onlinecite{du05}, from DFT PAW RPBE calculations.}
\footnotetext[8]{Ref.~\onlinecite{arboleda04} from PES calculations.}
\footnotetext[9]{Ref.~\onlinecite{sprunplu91} (see comments in the main text).}
\footnotetext[10]{Ref.~\onlinecite{johansson06} (see comments in the main text).}
\footnotetext[11]{This work.}
\end{table}

\subsection{H$_2$ dissociation and diffusion}\label{H2diss}

The activation barriers for H$_2$ dissociation over the various
metal-doped Mg surfaces are reported in Table~\ref{table2}, where we
also report two experimental values for the H$_2$
dissociation/recombination on the Mg(0001) surface. The value reported
in Ref.~\cite{sprunplu91} ($\sim$ 1.0 eV) refers to the recombination
barrier (which, in this particular case, is similar to the
dissociation barrier) identified with the barrier for desorption from
the surface.  This value was not directly measured in the thermal
programmed desorption (TPD) experiments of Ref.~\cite{sprunplu91}
because complete desorption spectra as function of temperature could
not be taken, due to the onset of Mg sublimation at $\sim$ 450 K which
overlaps with the temperature at which H$_2$ desorbs. However, it was
noted that the onset of H$_2$ desorption appears at 425 K, which is
similar to that of the H/Be(0001) system that has a determined
desorption energy of $\sim$ 1 eV~\cite{ray90}, and so, by analogy, it
was suggested that the activation energy for desorption might be the
same on the H/Mg(0001) system too.  The value reported in
Ref.~\cite{johansson06} (0.75$\pm$0.15 eV) has been obtained by the
interpretation of TPD experiments performed on a 400 \AA\, thick
magnesium film. This value is in good agreement with the calculated
PBE dissociation energy, but is significantly lower than the
dissociation energies calculated with RPBE by Vegge~\cite{vegge04} and
Du et al.~\cite{du05} which are 1.15 and 1.05 eV respectively. It
should be noted, however, as pointed out in Ref.~\cite{johansson06},
that the experimental situation may not be the same as the theoretical
ones, due to the possible presence of steps on the surface which might
be more reactive sites and lower the H$_2$ dissociation
barrier. Moreover, the inferred dissociation energy of 0.75$\pm$0.15
eV is based on the use of the Arrhenius relations with assumed
pre-factors of $\sim 10^{12}$ Hz.  As showed in Refs.~\cite{alfe06}
and~\cite{alfe07}, these values could be underestimated by more than
two orders of magnitudes because the classical pre-factors do not
include the enhancement due to the large entropy increase as the
molecules leave the surface, in which case the activation energy could
be up to $\sim 0.25$ eV higher.

The geometry of adsorption of the H atoms on the metal-doped Mg
surfaces listed in Table~\ref{table2} appears to be somewhat
correlated to the height of the dissociation barrier.  We find that
when the barrier is large (i.e., for Cu, Pd, Ag) the H atoms fall into
filled hollow sites, and when the barrier is null (i.e., for Ti, Zr,
V and Ru) they fall into empty hollow sites.  In between there are
elements showing a small energy barrier for which the preference
towards filled hollow sites (i.e., for Ni) instead of empty hollow
sites (i.e., for Fe, Co and Rh) is weaker (less than about 30 meV).

Fe is the only dopant for which magnetic calculations are really
required, with the total magnetic moment on the Fe atom being of 2.8
and 2.5 $\mu_B$ in the initial and final state of the dissociation
process respectively (Co is magnetic too, but when used as dopant of
the Mg surface our calculations show that it can be treated as
non-magnetic).  In particular, non-magnetic calculations for the
Fe-doped surface would give significantly different results, reducing
the dissociation barrier to almost zero and increasing by 60 \% the
energy difference between the initial and final states.

\begin{figure*}
\rotatebox{270}{\scalebox{0.9}[0.9]{\includegraphics{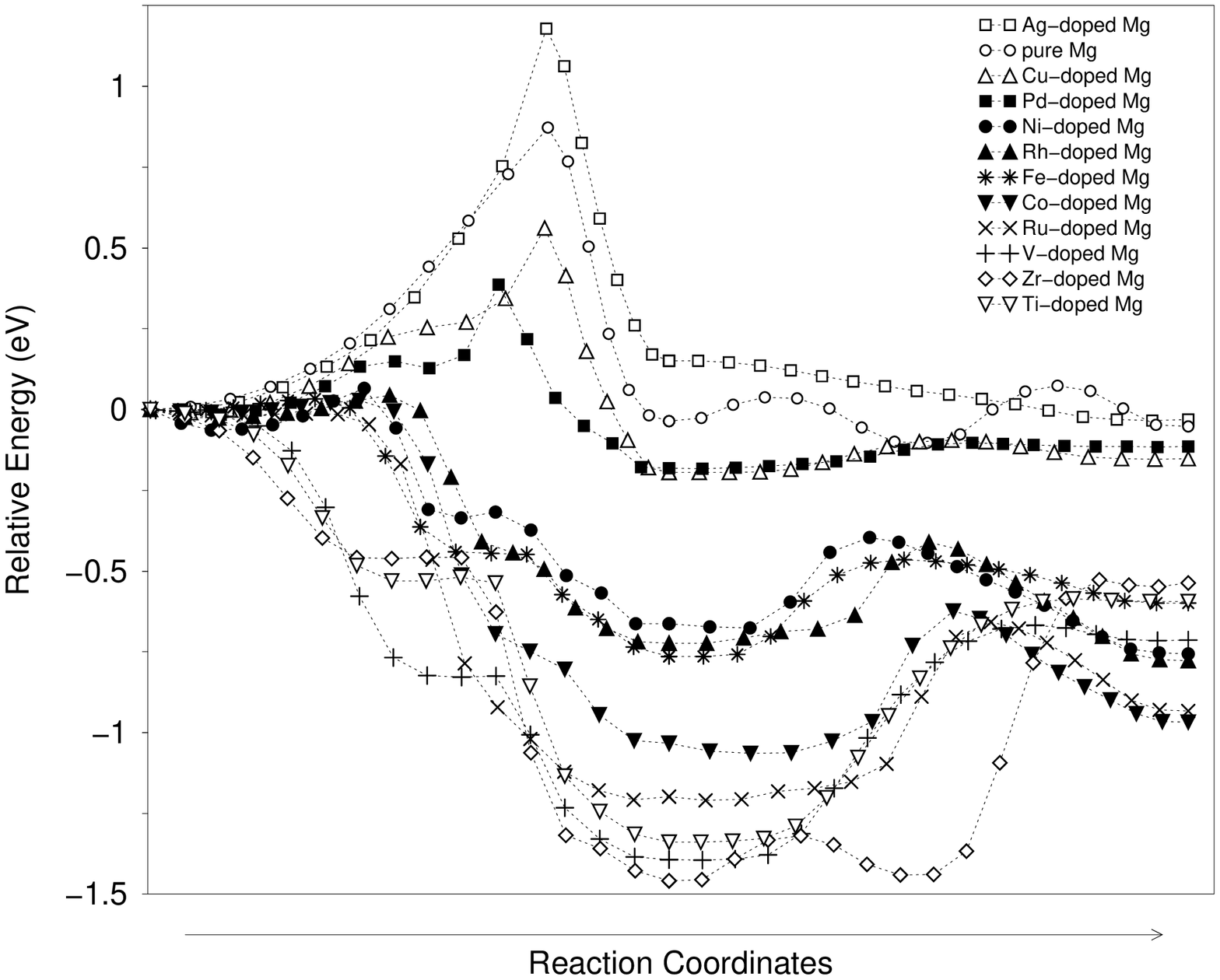}}}
\caption{\label{actbar}}
\phantom{xxxxxxxxxxxxxxxxxxxxxxxxxxxxxxxxxxxxxxxxxxxxxxxxxxxxxxxxxxxxxxxxxxxxxxxx} 
\phantom{xxxxxxxxxxxxxxxxxxxxxxxxxxxxxxxxxxxxxxxxxxxxxxxxxxxxxxxxxxxxxxxxxxxxxxxx}
\phantom{xxxxxxxxxxxxxxxxxxxxxxxxxxxxxxxxxxxxxxxxxxxxxxxxxxxxxxxxxxxxxxxxxxxxxxxx} 
\phantom{xxxxxxxxxxxxxxxxxxxx} 
\phantom{xxxxxxxxxxxxxxxxxxxx} 
\phantom{xxxxxxxxxxxxxxxxxxxx} 
\phantom{xxxxxxxxxxxxxxxxxxxx} 
\end{figure*}

As we noted before~\cite{pozzo}, the activation barrier for the
dissociation of H$_2$ over a metal-doped Mg surface is similar to that
on the corresponding pure metal surface.  For example, in the case of
Cu our calculated barrier is 0.56 eV, which is close to the
dissociation barrier of about 0.5 eV suggested by experiments and
other DFT (GGA and PBE)
calculations~\cite{hamnorNAT95,hammer94,gross94,murhog98,slham02,sakgross03}
over the pure Cu(111) surface. In addition, we find that the energy
difference between the final state and the initial state is -0.19 eV,
which is in line with the findings of Kratzer et al.~\cite{kratzer96},
who found that H$_2$ dissociation on the pure Cu(111) surface is
exothermic, with a gain of 0.2 eV.  On the Pd-doped surface we
calculate a dissociation barrier of 0.39 eV, which reduces to 0.30 eV
when a smaller 5x5x1 {\bf k}-points grid is used (instead of the 9x9x1
grid). The reason for mentioning the result obtained with the coarser
grid is because we want to compare with the findings of Dong, Kresse
\& Hafner~\cite{dong97}, who performed calculations with a similar
grid for the dissociation of H$_2$ on the Pd(111) surface, and found a
barrier of 0.29 eV when the molecule dissociates on top a Pd atom,
therefore a value very close to our value of 0.30 eV on the Pd-doped
Mg surface (note, however, that they report the bridge-bridge as the
preferred dissociation path, with a barrier of only 70 meV, which is
consistent with the theoretical value found by Nobuhara et
al.~\cite{nobuhara04} and in good agreement with the experimental
value of 50 meV~\cite{resch94}).  Finally, in the case of Ag, our
activation barrier of 1.18 eV is in agreement with the experimental
results which predict a dissociation barrier on Ag(111) larger than
that on Cu and somewhat larger than 0.8 eV~\cite{healey95}, and we
also agree with previous theoretical calculations~\cite{xu05} also
obtained on the pure Ag(111) surface, which reported an activation
energy of 1.11 eV.  Fig.~\ref{actbar} shows the MEP's for hydrogen
dissociation on a TM$=$ V, Fe, Co, Cu, Zr, Ru, Rh, Pd, Ag doped Mg
surface investigated here, together with those on a TM$=$Ni, Ti -doped
Mg surface that we have reported previously~\cite{pozzo}.

\begin{table*}
\caption{\label{table3} The d-band center position with respect to the
  Fermi energy ($E_d$), the activation energy barrier for the
  dissociation of H$_2$ (E$_{diss}$), the energy difference between
  the final and initial state (E$^{FS-IS}_{diss}$) of the
  dissociation, the activation energy barrier for the diffusion of
  atomic H (E$_{diff}$) and the corresponding energy difference
  (E$^{FS-IS}_{diff}$) on the pure Mg surface as opposed to the
  metal-doped Mg surfaces (these have been ordered so as to highlight
  the overall dependence along each column of the periodic table, as
  we go from right to left across the periodic table). Also reported
  in the last column is the average distance of molecular hydrogen
  from the surface as measured at the transition state,
  $\overline{d}_{H-surf}$.}
\begin{ruledtabular}
\begin{tabular}{lcccccc}
Surface & $E_d$ & E$_{diss}$ & E$^{FS-IS}_{diss}$ & E$_{diff}$ &
E$^{FS-IS}_{diff}$ & $\overline{d}_{H-surf}$ \\
 & (eV) & (eV) & (eV) & (eV) & (eV) & (\AA)  \\
\hline
Mg pure\footnotemark[1]      & --     & 0.87  & -0.04 & 0.11  & -0.02 & 0.9 \\ \\

Ag-doped Mg\footnotemark[2]  & -4.14  & 1.18  & 0.15  & null  & -0.18 & 0.7 \\ 
Cu-doped Mg\footnotemark[2]  & -2.27  & 0.56  & -0.19 & 0.10  &  0.04 & 0.7 \\ \\

Pd-doped Mg\footnotemark[2]  & -1.84  & 0.39  & -0.18 &  0.08 & 0.07  & 1.0 \\ 
Ni-doped Mg\footnotemark[1]  & -0.79  & 0.06  & -0.66 &  0.27 & -0.09 & 1.5 \\ \\

Rh-doped Mg\footnotemark[2]  & -0.75  & 0.04  & -0.72 &  0.31 & -0.05 & 1.8 \\ 
Co-doped Mg\footnotemark[2]  & -0.16  & 0.03  & -1.03 &  0.41 & 0.07 & 2.0 \\ \\

Ru-doped Mg\footnotemark[2]  & -0.14  & null  & -1.20 &  0.54 & 0.26 & -- \\ 
Fe-doped Mg\footnotemark[2]  & -0.72  & 0.03  & -0.76 &  0.30 & 0.17 & 2.0 \\ \\

V-doped Mg\footnotemark[2]   & +0.82  & null  & -1.39 &  0.73 & 0.68 & --  \\ \\

Zr-doped Mg\footnotemark[2]  & +1.32  & null  & -1.46 &  0.94 & 0.94 & -- \\ 
Ti-doped Mg\footnotemark[1]  & +1.08  & null  & -1.34 &  0.75 & 0.74 & -- \\
\end{tabular}
\end{ruledtabular}
\footnotetext[1]{Ref.~\onlinecite{pozzo}.}
\footnotetext[2]{This work.}
\end{table*}

\begin{figure*}
\rotatebox{0}{\scalebox{1.0}[1.0]{\includegraphics{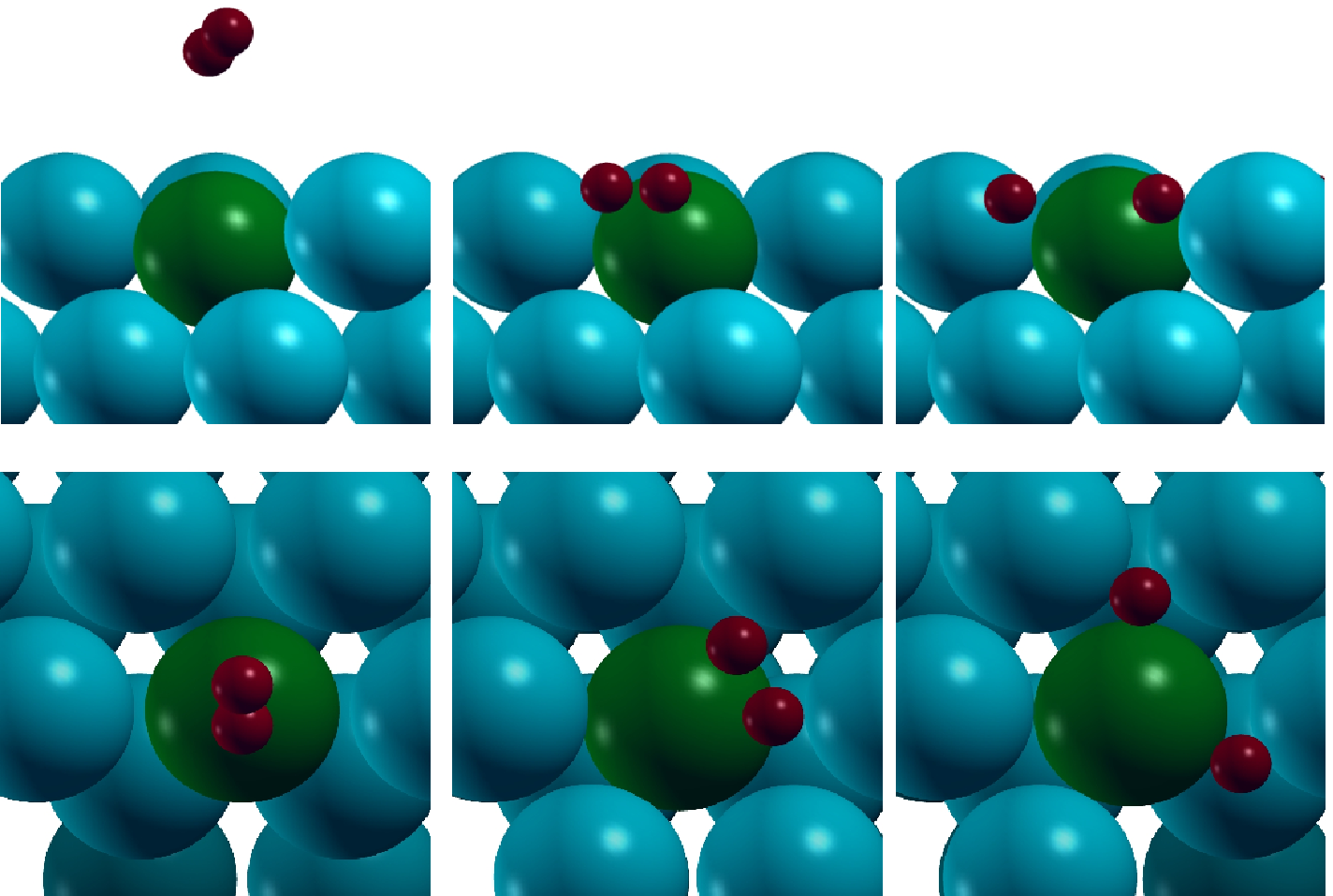}}}
\caption{\label{MgAgH2diss}}
\phantom{xxxxxxxxxxxxxxxxxxxxxxxxxxxxxxxxxxxxxxxxxxxxxxxxxxxxxxxxxxxxxxxxxxxxxxxx} 
\phantom{xxxxxxxxxxxxxxxxxxxxxxxxxxxxxxxxxxxxxxxxxxxxxxxxxxxxxxxxxxxxxxxxxxxxxxxx}
\phantom{xxxxxxxxxxxxxxxxxxxxxxxxxxxxxxxxxxxxxxxxxxxxxxxxxxxxxxxxxxxxxxxxxxxxxxxx} 
\phantom{xxxxxxxxxxxxxxxxxxxx} 
\phantom{xxxxxxxxxxxxxxxxxxxx} 
\phantom{xxxxxxxxxxxxxxxxxxxx} 
\phantom{xxxxxxxxxxxxxxxxxxxx} 
\end{figure*}

Figure~\ref{MgAgH2diss} shows the dissociation of H$_2$ on the
Ag-doped Mg surface, which is found to have the largest barrier value
among all the dopants investigated here (since this barrier is larger
than that on pure Mg, obviously H$_2$ will not dissociate onto this
site, but will rather choose some regions of the Mg surface free of
Ag).  For Fe, Co, Cu, Rh and Pd doped Mg surfaces, the images at the
IS, TS and FS of the NEB are similar, with the H$_2$ molecule at the
TS sitting closer (i.e., Cu and Pd) or further away from the surface
(i.e., Fe, Co and Rh) compared to the behavior shown in our previous
paper~\cite{pozzo} for the Ni-doped Mg surface (see
$\overline{d}_{H-surf}$ values reported in Table~\ref{table3}).  For
Ag we note that the hydrogen molecule at the TS dissociates closer to
the surface than on the Ni-doped Mg surface, and that it does so on a
side of the dopant atom (see Figure~\ref{MgAgH2diss}).

A closer look at the geometry of the dissociation process shows an
interesting correlation between the height of the barrier and the
geometry of the transition state. The dissociation of molecular
hydrogen into two hydrogen atoms happens on-top of the dopant atom
when Fe, Co, Ni or Rh are used as dopants, slightly shifted to the
side when Pd and Cu are the metals used as dopants, and fully on the
side when Ag is the dopant.  In other words, it appears that H$_2$
dissociates on top of the dopant atom for those doped Mg surfaces
which show a very small barrier (i.e., Fe, Co, Ni and Rh), slightly
shifted to the side of the dopant atom for the Pd and Cu doped Mg
surfaces having a non negligible barrier, and completely on the side
of the dopant atom on the Ag-doped Mg surface which shows a very large
dissociation barrier.

The dissociation of the H$_2$ molecule is only the first step for the
absorption of hydrogen. A second fundamental step is the diffusion of
the products away from the catalytic site. To study this, we performed
NEB calculations in which the initial state was the final state of the
dissociation process, and the final state was obtained by displacing
one H into a nearby hollow site.  Figure~\ref{MgFeHdiff} shows the
diffusion path of one of the two hydrogen atoms on the Fe-doped Mg
surface as an example. The MEP's for the diffusion processes are also
shown in Fig.~\ref{actbar}.  We observe that the height of the
diffusion barrier E$_{diff}$ is strongly anti-correlated to the height
of the dissociation barrier E$_{diss}$ (see Table~\ref{table3}). In
fact, Ti, V, Zr and Ru have zero dissociation barriers, but they bind
the products very strongly, which results in high values of
E$_{diff}$. By contrast, Ag, Cu and Pd produce large dissociation
barriers, but they have low diffusion barriers (in fact, no barrier at
all for Ag). In between there are Fe, Ni and Rh, which
represent the best compromise in combining low activation barriers for
both processes. Ni is the best possible choice overall.

We note in passing that the catalytic effect of Ni dopant on MgH$_2$
for the dehydrogenation process (not studied here) has been
experimentally demonstrated by Jensen et al.~\cite{jensen06}, showing
an activation energy reduced by 0.5 eV with respect to that obtained
with pure MgH$_2$.

\begin{figure}
\rotatebox{-90}{\scalebox{0.65}[0.68]{\includegraphics{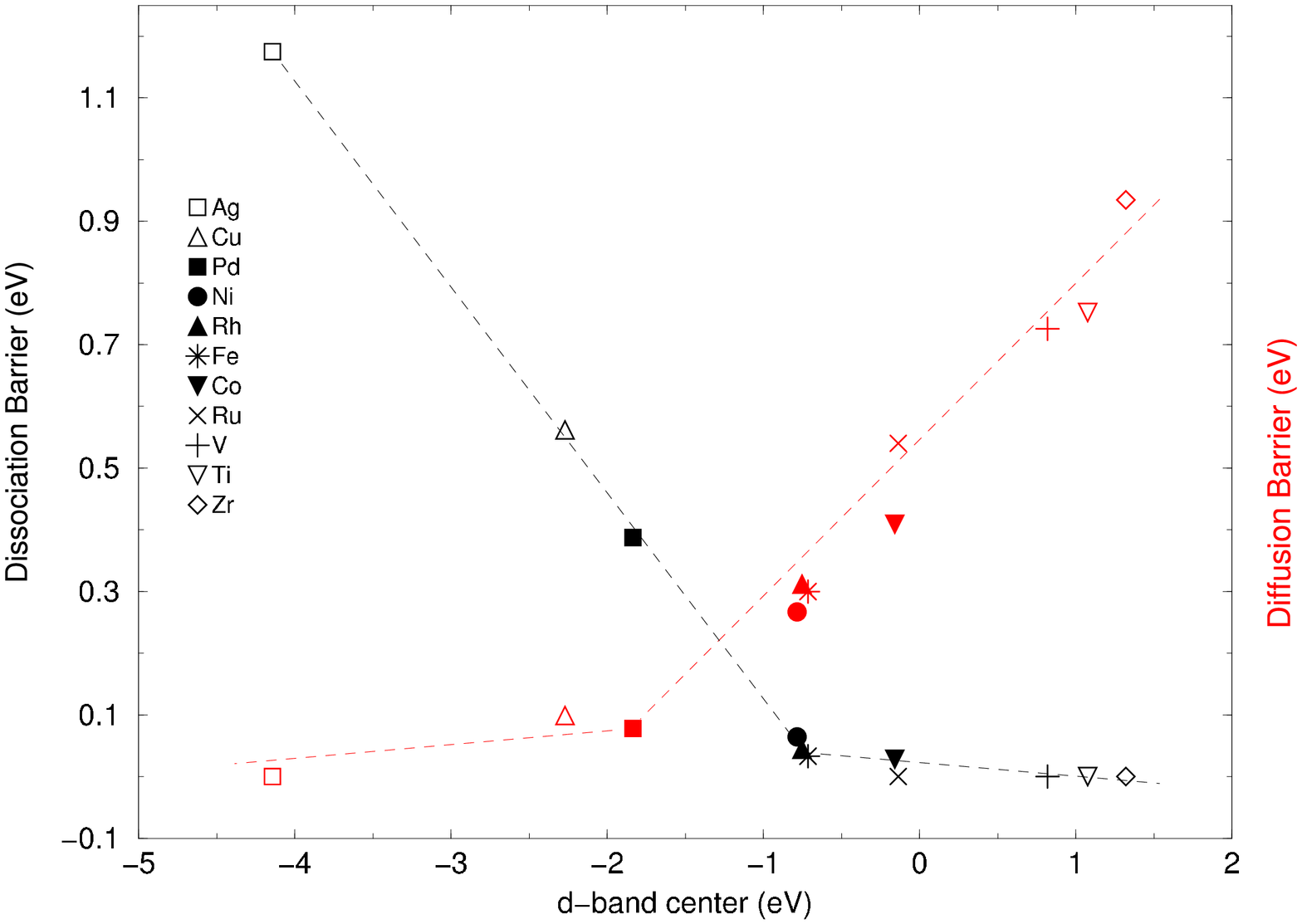}}}
\caption{\label{db_actbar}}
\phantom{xxxxxxxxxxxxxxxxxxxxxxxxxxxxxxxxxxxxxxxxxxxxxxxxxxxxxxxxxxxxxxxxxxxxxxxx} 
\phantom{xxxxxxxxxxxxxxxxxxxxxxxxxxxxxxxxxxxxxxxxxxxxxxxxxxxxxxxxxxxxxxxxxxxxxxxx}
\phantom{xxxxxxxxxxxxxxxxxxxxxxxxxxxxxxxxxxxxxxxxxxxxxxxxxxxxxxxxxxxxxxxxxxxxxxxx} 
\phantom{xxxxxxxxxxxxxxxxxxxx} 
\phantom{xxxxxxxxxxxxxxxxxxxx} 
\phantom{xxxxxxxxxxxxxxxxxxxx} 
\phantom{xxxxxxxxxxxxxxxxxxxx} 
\end{figure}

It is interesting to correlate the height of the barriers with the
position of the $d$-band of the transition metal dopant with respect
to the Fermi energy $E_F$ (here we define the $d$-band, $p_d(E)$, as
the projection of the electronic density of states onto $d$ type
spherical harmonics). In particular, it is useful to consider the
first energy moment of the $d$-band, or $d$-band centre, defined as
$E_d = \int_{-\infty}^{E_0}dE(E - E_F)p_d(E)$, where $E_0$ is some
cutoff energy which we chose to be 7 eV above the Fermi energy.  In
Fig.~\ref{db_actbar} we plot the dissociation and the diffusion energy
barriers, E$_{diss}$ and E$_{diff}$, against $E_d$ for all the systems
explored (see Table~\ref{table3}).  It is obvious that the heights of
the barriers are strongly correlated with the position of the $d$-band
centre.

The step limiting process in the hydrogen absorption is the one with
the largest energy barrier between dissociation and diffusion, so the
best dopant is the one which minimizes the largest energy barrier. It
is customary to define the activity of a catalyst in terms of the rate
of the reaction which is being catalysed. This can often be accurately
approximated by an Arrhenius relation, and therefore the natural
logarithm of the rate is proportional to the negative of the
activation energy barrier. We can then interpret the maximum of the
two barriers shown in Fig.~\ref{db_actbar} as indicating the activity
of the catalyst. If we draw a line across these points, we see that
the various transition metals investigated here fit on an inverse {\em
  volcano plot}, with Ni, Fe and Rh sitting near the top of the
volcano, and therefore being the most active catalysts.

\begin{figure}
\rotatebox{-90}{\scalebox{0.65}[0.68]{\includegraphics{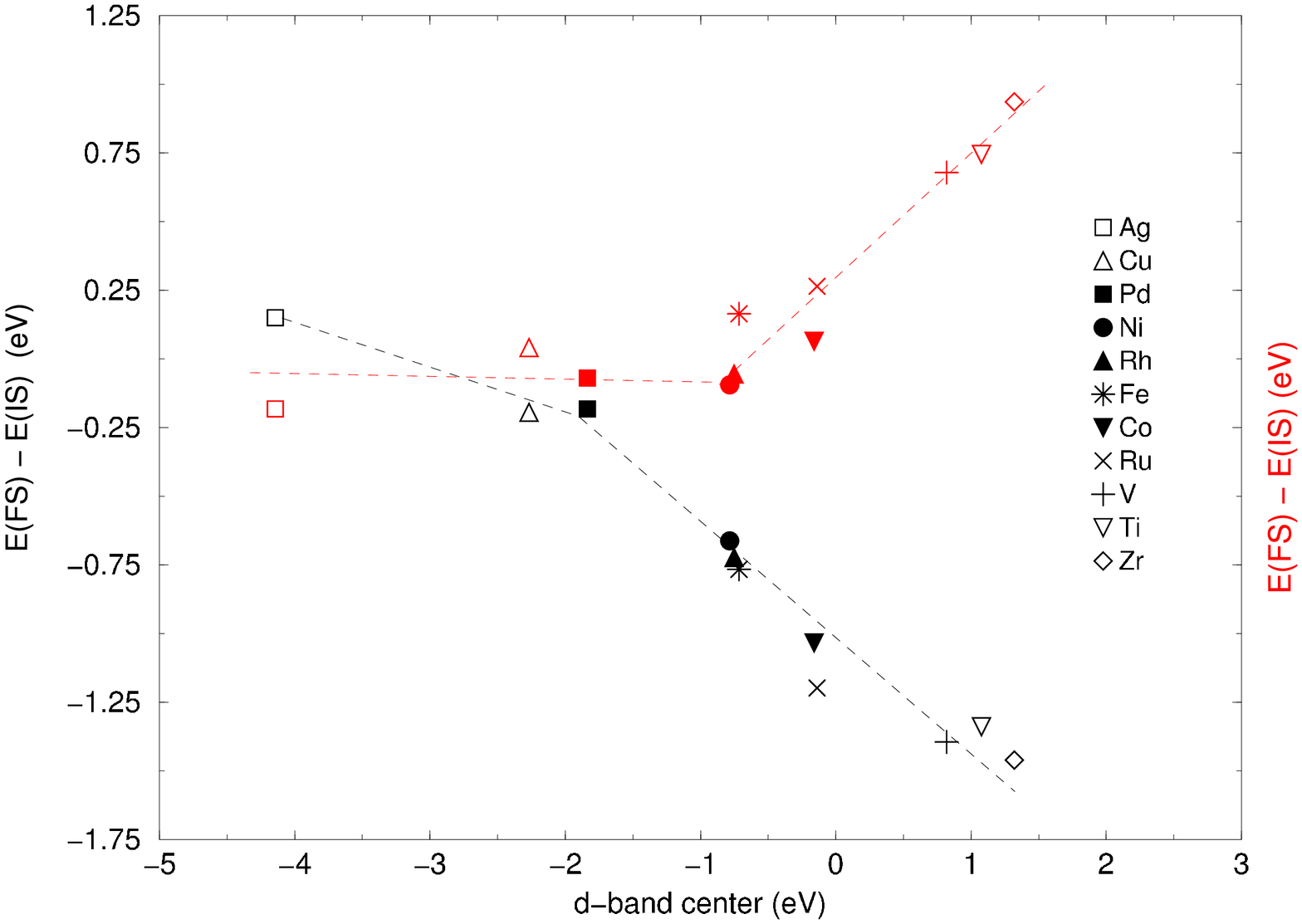}}}
\caption{\label{db_isfs}}
\phantom{xxxxxxxxxxxxxxxxxxxxxxxxxxxxxxxxxxxxxxxxxxxxxxxxxxxxxxxxxxxxxxxxxxxxxxxx} 
\phantom{xxxxxxxxxxxxxxxxxxxxxxxxxxxxxxxxxxxxxxxxxxxxxxxxxxxxxxxxxxxxxxxxxxxxxxxx}
\phantom{xxxxxxxxxxxxxxxxxxxxxxxxxxxxxxxxxxxxxxxxxxxxxxxxxxxxxxxxxxxxxxxxxxxxxxxx} 
\phantom{xxxxxxxxxxxxxxxxxxxx} 
\phantom{xxxxxxxxxxxxxxxxxxxx} 
\phantom{xxxxxxxxxxxxxxxxxxxx} 
\phantom{xxxxxxxxxxxxxxxxxxxx} 
\end{figure}

As a matter of interest, in Fig.~\ref{db_isfs} we plot the energy
difference between the final and the initial state $E^{FS-IS}$ (both
for the dissociation and for the diffusion process) as a function of
the $d$-band centre. We can observe some correlation between the two
quantities, although this is less strong than that observed in
Fig.~\ref{db_actbar} for the height of the two energy barriers. If
follows that the correlation between the energy barriers and
$E^{FS-IS}$ is also weaker than that between the energy barriers and
the $d$-band centre, leaving the latter a better parameter to
characterize the catalyst.

From an inspection of Figs.~\ref{db_actbar} and ~\ref{db_isfs} the
$d$-band centre correlation is evident, and points to an ideal
$d$-band centre value of about -1.29 eV.  This value cannot be
obtained with any of the TM-doped Mg surfaces investigated
here. Recently, Vegge et al.~\cite{vegge05} have investigated
magnesium 3d TM alloys. They showed that the $d$-band centre values of
the expanded alloys obtained with TM belonging to the first raw of the
Periodic Table range from +0.93 to -6.88 eV going from MgSc to
MgZn. In particular, MgCu gives a value of -2.37 eV while the neighbor
MgNi -0.82 eV. It would be interesting to broaden their investigation
to 4$d$ and 5$d$ TMs to see if the optimal $d$-band centre value of
about -1.3 eV that we have extrapolated here could be obtained with
some alloys, but this is beyond the purpose of the present
investigation. \\ \\

\section{Conclusions}\label{conclusions}

We have performed here a systematic DFT/PBE study of hydrogen
dissociation and subsequent diffusion over Mg surfaces doped with
different transition metals. The dopants investigated were Ti, Zr, V,
Fe, Ru, Co, Rh, Ni, Pd, Cu and Ag. We have observed that the
transition metals on the left of the periodic table (Ti, V, Zr),
together with Ru, eliminate the dissociation barrier altogether,
however, the products stick too strongly to the metal dopant,
therefore hindering diffusion away from the catalytic site. This would
result in a quick deactivation of the catalyst and therefore a slow
absorption process.  On the contrary, the transition metals on the
right of the periodic table do not bind too strongly the H atoms (in
fact, Ag does not bind them at all), allowing easy diffusion, however,
their effect on the dissociation barrier is small.  We have shown that
these two opposite catalytic properties are well correlated to the
position of the $d$-band centre, according to the Hammer \&
N{\o}rskov~\cite{hamnor95} model.  In fact, we have shown that the
catalytic activity for the H absorption process can be described well
by a volcano plot, with the most active catalysts Ni, Fe and Rh
sitting near the top of the volcano.

\begin{acknowledgments}
This work was conducted as part of a EURYI scheme award as provided by EPSRC 
(see www.esf.org.euryi). Calculations have been performed on the LCN cluster 
at University College London. We thank Sam French and Alvaro Amieiro for very
useful discussions.
\end{acknowledgments}

\newpage

\noindent{\bf  List of Figures} 

\begin{enumerate} 

\item{Minimum energy paths for the dissociation of the H$_2$ molecule, and 
subsequent diffusion of one of the two H atoms, on a pure Mg surface and on 
(Ti, V, Fe, Co, Ni, Cu, Zr, Ru, Rh, Pd, Ag)-doped Mg surfaces.}

\item{(Colour online) H$_2$ (dark red) dissociation over the Ag-doped 
Mg surface as viewed from side (top figures) and top (bottom figures) positions 
respectively at IS (left-hand panel), TS (central panel) and FS (right-hand panel). 
The Mg, Ag and H atoms are represented respectively 
by light grey, dark grey and black colours.}

\item{(Colour online) H (dark red) diffusion on the Fe-doped Mg surface as viewed 
from top. Figures show positions at the final state of the dissociation which is 
the initial state for the diffusion process (left), at the transition state (centre) 
and final state (right) of the diffusion process. The Mg, Fe and H atoms are 
represented respectively by light grey, dark grey and black colours.}

\item{Activation energy barrier for hydrogen dissociation (black) and 
diffusion (red) of hydrogen on pure Mg and metal-doped Mg surfaces as a function of 
the $d$-band center positions. The dashed lines have been drown for eye guidance only.}

\item{The energy difference between the final and initial state, E(FS-IS), 
for hydrogen dissociation (black) and diffusion (red) on pure Mg and metal-doped Mg surfaces 
as a function of the $d$-band center positions. The dashed lines have been drown for eye 
guidance only.}

\end{enumerate}

\newpage

\newpage

\newpage

\newpage

\end{document}